\documentstyle[12pt]{article} \tolerance=10000
\begin{document}
\begin{flushright}
hep-th/0209136\\
SNBNCBS-2002
\end{flushright}
\vskip 3.7cm
\begin{center}
{\bf \Large { Abelian 2-form gauge theory: special features}}

\vskip 2cm

{\bf R. P. Malik}
\footnote{ E-mail: malik@boson.bose.res.in  }\\
{\it S. N. Bose National Centre for Basic Sciences,} \\
{\it Block-JD, Sector-III, Salt Lake, Calcutta-700 098, India} \\

\vskip 2.5cm

\end{center}
{\bf Abstract:}               
It is shown that the four $(3 + 1)$-dimensional (4D) free Abelian 
2-form gauge theory provides an example of (i) a class of field theoretical
models for the Hodge theory, and
(ii) a possible candidate for the quasi-topological
field theory (q-TFT). Despite many striking similarities with some
of the key topological features
of the two $(1 + 1)$-dimensional (2D) free Abelian (and self-interacting
non-Abelian) gauge theories, it turns out that the 4D free Abelian 2-form
gauge theory is {\it not} an exact TFT. To corroborate this
conclusion, some of the key issues are discussed. In particular, it is shown 
that the (anti-)BRST and (anti-)co-BRST invariant quantities of the
4D 2-form Abelian gauge theory obey the
recursion relations that are reminiscent of the exact TFTs
but the Lagrangian density of this theory is not found to be able to 
be expressed as the sum
of (anti-)BRST and (anti-)co-BRST exact quantities as is the case with
the {\it topological} 2D free Abelian (and self-interacting non-Abelian) 
gauge theories.\\

\baselineskip=16pt

\vskip 1cm

\newpage

\noindent
{\bf 1 Introduction}\\

\noindent
In recent years, there has been an upsurge of interest in the study 
of some of the key physical and mathematical
issues associated with the 4D massless
2-form (antisymmetric) Abelian gauge theory which has already been proven to
provide a dual description for the massless scalar field theory 
(see, e.g. [1-3]). Such an interest has been thriving 
for the past few years due to the relevance of the basic field 
(a second rank antisymmetric tensor $B_{\mu\nu} = - B_{\nu\mu}$)
of this theory in the context of modern developments in the subject of
(super)string theories, related issues of the extended objects,
supergravity  theories and non-commutative field theories. For instance,
the antisymmetric potential $B_{\mu\nu}$ of this 2-form gauge theory 
appears very naturally in the supergravity multiplets
[4] and in the excited states of the quantized (super)string theories
(see, e.g., [5,6] for details). The existence of this field is also crucial
to the anomaly cancellation mechanism for the superstring theories and its
presence provides an estimate for the dualities in the extended objects 
(see, e.g., [5,6]).
This field has also turned up in an elegant way in the context of modern
developments in the subject of non-commutative geometry [7]. 
Besides the above relevance and importance, the 
antisymmetric potential and the corresponding gauge theory have
been an interesting topic of discussion for diverse and distinct 
reasons in various branches of
theoretical physics, e.g., cosmic string theory, vortices in an incompressible
and irrotational fluid, QCD, 
``hairs'' on the black holes, etc. [8-11]. It is now quite well known
that this 2-form antisymmetric $B_{\mu\nu}$
field generates an effective mass for
the one-form Abelian gauge field $A_\mu$ in 4D through a topological coupling 
(i.e. celebrated $B \wedge F$) term
where the gauge invariance and mass co-exist 
{\it together} without the presence of 
a residual scalar Higgs field [3,11,12]. This latter theory has been 
studied from various points of view because of
its innate rich mathematical structure and physical
relevance. In this connection, mention can be
made of interesting discussions related to this theory
such as Dirac bracket analyses [12-14], BFT Hamiltonian
formulations [15], BRST quantization [16], duality considerations [17,18], etc.

The covariant canonical quantization of the massless 2-form gauge theory
was first attempted in [19]. This aspect of quantization
 has been subsequently studied
in a systematic manner by a host of authors in the framework of
Becchi-Rouet-Stora-Tyutin (BRST) formalism [20-24]. It was observed that this 
theory possesses  an interesting constraint structure and a naive 
gauge-fixing term (containing the antisymmetric tensor field) is found to
be invariant under a secondary gauge transformation. This intrinsic
property of the theory requires the introduction of ghosts for ghost fields 
so that the gauge-fixing can be achieved in a complete and
systematic fashion for the 
purpose of BRST quantization. In all the above approaches [19-24], the 
generator of the BRST transformation (i.e. the nilpotent ($Q_{b}^2 = 0$)
and conserved ($\dot Q_{b} = 0$)
BRST charge $Q_{b}$) turns out to define the BRST cohomology in the
quantum Hilbert space of states. In the language of the
differential geometry and cohomology, the analogue of the exterior 
derivative $d$ (with $d^2 = 0, d =  dx^\mu \partial_\mu)$ is none other than 
the BRST charge $Q_{b}$ connected with the 
(gauge) description of the 2-form massless gauge theory. In a recent paper
[25], however, the usual BRST formalism with a single BRST charge $Q_{b}$ (i.e.
the analogue of $d$) has been extended to include the symmetry
generators (as conserved charges)
$Q_b, Q_d, Q_w$ that produce local, covariant and continuous 
symmetry transformations for the Lagrangian density of the theory. The
latter two symmetry transformations, 
generated by $Q_d$ and $Q_w$, correspond to the co-exterior derivative
$\delta = \pm * d *$ (with $\delta^2 = 0$) and the Laplacian operator
$\Delta = (d + \delta)^2 = d \delta + \delta d$
where $*$ is the Hodge duality operation [26,27].  We christen the symmetry
transformations on the fields corresponding to the
analogue of $\delta$ as the dual(co)-BRST symmetry transformations
and a bosonic symmetry (which is equal to the anti-commutators of
nilpotent (anti-)BRST and (anti-)co-BRST symmetries) corresponds to
the Laplacian operator $\Delta$. One can explain the existence of these 
symmetries in the language of the invariances of certain specific terms of the
Lagrangian density of the theory which are closely connected with the
cohomological operators. For instance, the (anti-)BRST transformations
keep the kinetic energy term (or, more precisely, the curvature term $H = d B$
itself
\footnote{ The antisymmetric gauge potential $B_{\mu\nu}$ is defined
through the 2-form $B = \frac{1}{2!} B_{\mu\nu} dx^\mu \wedge dx^\nu$.
The curvature term $H_{\mu\nu\kappa}$ is defined through the application of $d$
on $B$ (i.e. $ H = d B, H = \frac{1}{3!} H_{\mu\nu\kappa} dx^\mu \wedge dx^\nu
\wedge dx^{\kappa})$. The gauge-fixing term
$(\partial_\mu B^{\mu\nu})$, on the other hand,
is defined through the application of $\delta$
on $B$ (i.e. $ \delta B = - * d * B = \partial_\mu B^{\mu\nu} d x_{\nu}$).
Thus, in some sense, the gauge-fixing term is `Hodge dual' to the curvature
term because $d$ and $\delta = - * d *$ are Hodge dual to each-other
(see, e.g., [26-30] for details).}) 
invariant. In contrast, it is the gauge-fixing term 
($\delta B = \partial_\mu B^{\mu\nu} dx_\nu$) that remains 
invariant under the (anti-)co-BRST transformations. The bosonic symmetry 
corresponding to the Laplacian operator leaves the ghost fields invariant
(or ghost fields, at the most, transform by a vector gauge transformations).
Together all the three symmetry transformations of the theory  correspond 
to all the de Rham cohomological operators $(d,\delta,\Delta)$ of the 
differential geometry so that the analogue of the Hodge decomposition theorem 
(HDT) could be defined in the quantum Hilbert space of states in terms of 
the above conserved charges [25]. There exists a discrete symmetry in 
the theory (cf (2.4) below) that corresponds to the Hodge duality 
$*$ operation of the differential geometry [25].

It would be nice to recall that, analogous to the
4D 2-form Abelian gauge theory,
the free 2D Abelian $U(1)$ gauge theory [31-33,41,43], the 2D self-interacting
non-Abelian gauge theory (where there is no interaction between gauge fields 
and matter fields) [33,34,41,42] and the 2D interacting $U(1)$ gauge theory
(where there is an interaction between the $U(1)$
gauge field and Dirac (matter) fields)
[35,36] have been shown to provide a set of field theoretical models for the
Hodge theory where the local, covariant and continuous symmetry transformations
(and their generators) correspond to the three de Rham cohomological operators
($d,\delta, \Delta$) of differential geometry. A discrete symmetry has been 
shown to exist for all the above theories which corresponds to the Hodge
duality $*$ operation of the differential geometry. Furthermore, the 2D free
Abelian and self-interacting non-Abelian gauge theories have been shown to be 
topological
\footnote { A (gauge) theory defined on the flat spacetime manifold 
(with a flat spacetime metric) and
having no propagating degrees of freedom associated with the basic (gauge)
field of the theory (see, e.g., [37]).}
in nature because of the existence of the topological invariants
(as well as their proper recursion relations) and the
appearance of the Lagrangian density and symmetric energy momentum tensor 
which could be expressed as the sum of the (anti-)BRST and (anti-)co-BRST
anti-commutators [33,38-41].
The geometrical interpretations for all the symmetry transformations and
corresponding conserved charges (as generators) have been discussed
in the framework of superfield formulation 
(see, e.g., [38-43] for details). In a recent set of papers [39-43], 
for the first time, the geometrical origin for the 
topological and cohomological
aspects of these theories have been provided in the framework of
superfield formulation. Thus, it will be noted that
the above cited theories have very rich mathematical structure associated
with them as they belong to a new class of topological field theories (TFTs)
where the appearance of the Lagrangian density 
(and the symmetric energy momentum tensor)
is like the Witten type TFTs [44]
but the symmetries of these theories are just like Schwarz type 
(without a topological shift symmetry which is a characteristic feature of
the existence of a Witten type TFT) [45].

It is obvious from the above discussions that the 4D 2-form Abelian
gauge theory and 2D free Abelian and self-interacting non-Abelian gauge 
theories are prototype examples of a class of field theoretical models for
the Hodge theory. The latter 2D theories are also {\it exact} 
TFTs because, the
existence of (anti-)BRST and (anti-)co-BRST symmetries in the theory,
turns out to be responsible for gauging out both the propagating
degrees of freedom of 2D photon (and/or 2D gluon) (see, e.g.,
 [31-33,37] for details). In the recent past,
there has been some interest in studying {\it almost} TFTs (or quasi-TFTs)
which involve constraints that {\it leave out only a finite 
number of degrees of freedom in the theory} [37,46]. So far, the 
(quasi-)topological and cohomological nature of the 4D 2-form
gauge theory have not been studied together in
a complementary fashion. The purpose of the present paper is 
to study the 4D 2-form Abelian gauge theory from the point of view of 
gaining some insights into the theoretical structure behind its 
quasi-topological as well as cohomological nature. We are propelled to
go for such studies because of the close kinship that exists
between this theory and the
2D free (and self-interacting) (non-)Abelian gauge theories. Our study is
essential mainly on four counts: (i) any new property associated with
the 4D Abelain 2-form gauge theory will turn out to be useful in the
context of modern developments in (super)string theories and related
topological and/or solitonic extended objects. (ii) The present theory
is the {\it first} example of a $(3 + 1)$-dimensional gauge theory that 
provides a field theoretical model for the Hodge theory. It would be 
interesting, from the cohomological as well as topological points of view,
to study this theory in more detail so that the idea of HDT and its connection 
with the (quasi-)TFTs can be put on a firmer footing. Specially, 
one should look for
the quasi-topological nature of this theory as there is already a scalar
degree of freedom associated with $B_{\mu\nu}$ in 4D. (iii) The study of
this theory on 4D manifold with flat spacetime metric is the  first step
in the direction to our main goal of
studying this theory on the curved spacetime manifold
with a non-trivial spacetime dependent metric. (iv) Our present study 
{\it might} turn out to be useful in the study of
an ``interacting'' theory where $B \wedge F$
term is present. It might also give some insights into the way to attempt
the study of non-Abelian 2-form gauge theory in the ``extended''
framework of BRST formulation.

In our present paper, we shall concentrate mainly on the
(anti-)BRST and (anti-)co-BRST invariant versions (cf (2.1) and (2.8) below)
of the following $(3 + 1)$-dimensional
\footnote{ We adopt here the conventions and
notations in which the 4D flat Minkowski metric is:
$\eta_{\mu\nu} = $ diag$\; (+1, -1, -1, -1)$  
and totally anti-symmetric Levi-Civita tensor 
($\varepsilon_{\mu\nu\kappa\zeta}$) is chosen such that:
$\varepsilon_{0123} = + 1 = - \varepsilon^{0123}, \varepsilon_{0ijk} =
\varepsilon_{ijk} = - \varepsilon^{ijk}$,
$\varepsilon_{\mu\nu\kappa\zeta} \varepsilon^{\mu\nu \kappa\zeta} = - 4!,
\varepsilon_{\mu\nu\kappa\zeta} \varepsilon^{\mu\nu\kappa\eta} = - 3!
\delta_\zeta^{\eta}, \varepsilon_{\mu\nu\kappa\zeta} 
\varepsilon^{\mu\nu\eta\sigma} = - 2! 
(\delta^\eta_\kappa \delta^\sigma_\zeta
- \delta^\eta_\zeta \delta^\sigma_\kappa)$ etc.
Here Greek indices: 
$\mu, \nu, \kappa.. = 0, 1, 2, 3$ stand for the 
spacetime directions on the manifold and Latin indices:
$i, j, k...= 1, 2, 3$ correspond to the space directions only.}
free Kalb-Ramond Lagrangian density [1,2,8]
$$
\begin{array}{lcl}
{\cal L}_{0} =  \frac{1}{12}\; H^{\mu\nu\kappa} H_{\mu\nu\kappa}
\equiv 
 \frac{1}{2} {\cal B}_\mu {\cal B}^\mu
- \frac{1}{3!} \varepsilon_{\mu\nu\kappa\zeta} {\cal B}^\mu 
H^{\nu\kappa\zeta}
\end{array}\eqno(1.1)
$$
where $H_{\mu\nu\kappa} = \partial_\mu B_{\nu\kappa} + \partial_\nu 
B_{\kappa\mu} + \partial_\kappa B_{\mu\nu}$
is the curvature tensor constructed from the antisymmetric
tensor field $B_{\mu\nu}$ and the vector auxiliary field ${\cal B}_\mu$
has been introduced to linearize the quadratic kinetic energy term
$(\frac{1}{12}\; H^{\mu\nu\kappa} H_{\mu\nu\kappa})$. 
It is evident that, in the above, ${\cal B}_\mu$ and
$H_{\mu\nu\kappa}$ have the following inter-relationships due to 
the Euler-Lagrange equation of motion
$$
\begin{array}{lcl}
{\cal B}_\mu = \frac{1}{3!}\; \varepsilon_{\mu\nu\kappa\zeta}
 H^{\nu\kappa\zeta}\;
\equiv \;\frac{1}{2}\; \varepsilon_{\mu\nu\kappa\zeta} \partial^\nu 
B^{\kappa\zeta}\; \qquad \;H_{\mu\nu\kappa} = \varepsilon_{\mu\nu\kappa\zeta}
{\cal B}^\zeta.
\end{array}\eqno(1.2)
$$
We shall demonstrate, in our present endeavour,
that (i) the off-shell nilpotent (anti-)BRST and (anti-)co-BRST invariant
Lagrangian density (2.1) can be expressed as the sum of: (a) the kinetic energy
term and an (anti-)BRST exact part and/or (b) the gauge-fixing term and
an (anti-)co-BRST exact part. This feature is similar to the 2D free
Abelian and self-interacting non-Abelian  gauge theories which are 
{\it topological} in nature [33,39,40]. 
(ii) There exist twelve sets of (anti-)BRST and (anti-)co-BRST
invariant quantities for the Lagrangian density (2.1) that obey the
recursion relations that are reminiscent of such relationships
in the context of exact TFTs. (iii) The Lagrangian
density (2.8) could be expressed, in terms of the {\it on-shell} nilpotent
(anti-)BRST and (anti-)co-BRST charges, as the sum of (anti-)BRST and
(anti-)co-BRST anti-commutators {\it but for a factor of half}
(cf eqn (4.14) below). (iv) The
4D free Abelian 2-form gauge theory is very likely a candidate for the
field theoretical model for a q-TFT and, of course, it is a very beautiful 
as well as an exact field theoretical model for the Hodge theory [25].

The outline of our paper is as follows. In section 2, we set up the notations
as well as conventions by recapitulating some of the salient features
of our earlier work [25] on the existence of off-shell nilpotent (anti-)BRST 
and (anti-)co-BRST symmetries in the Lagrangian formulation. 
We also discuss briefly the on-shell nilpotent version of these symmetries 
for our later convenience. Section 3 is central of our present paper and is
devoted to the discussion of some special features
associated with the 4D 2-form Abelian gauge theory. 
These features are mainly (i) the proof
of this theory to be a field theoretical model for the
Hodge theory [25], and (ii) its quasi-topological nature. Both these issues 
are so elegantly intertwined with each-other that first we present some of the
key cohomological properties [25] and, then only, 
we discuss the quasi-topological nature.
We compare and contrast some of the decisive properties
of the 4D 2-form and 2D one-form Abelian gauge theories in section 4. Finally,
in section 5, we make some concluding remarks.\\

\noindent
{\bf 2 Preliminary: (co-)BRST symmetries}\\

\noindent
We briefly recapitulate here
the off-shell- as well as on-shell nilpotent (anti-)BRST  
and (anti-)dual-BRST symmetries 
which would be useful for our later discussions.\\

\noindent
{\bf 2.1 Off-shell nilpotent symmetries}\\

\noindent
We provide here a concise synopsis of our earlier work [25]
on the existence of
the off-shell nilpotent ($s_{(A)B}^2 = s_{(A)D}^2 = 0$) (anti-)BRST
($s_{(A)B}$) and (anti-)dual-BRST ($s_{(A)D}$) symmetries that are
respected {\it together} by the following linearized version of the
Lagrangian density for the 2-form free Abelian 
gauge theory in $(3 + 1)$-dimensions [25] 
$$
\begin{array}{lcl}
{\cal L}^{(l)}_{B} &=& 
 \frac{1}{2}\; {\cal B}^\mu {\cal B}_\mu 
- {\cal B}^\mu (\frac{1}{2}\; 
\varepsilon_{\mu\nu\kappa\zeta} \partial^\nu B^{\kappa\zeta} 
- \partial_\mu \phi_{2})
+ B^\mu (\partial^\nu B_{\nu \mu} - \partial_\mu \phi_{1})
- \frac{1}{2}\; B^\mu B_\mu 
\nonumber\\
&-& \partial_\mu \bar \beta \;\partial^\mu \beta
+ (\partial_\mu \bar C_\nu - \partial_\nu \bar C_\mu) (\partial^\mu C^\nu)
+ \rho \;(\partial_\mu C^\mu + \lambda) + (\partial_\mu \bar C^\mu + \rho)\;
\lambda
\end{array}\eqno(2.1)
$$
where $B_\mu, {\cal B}_\mu, \phi_{1}, \phi_{2}$ are the bosonic auxiliary
fields and $B_{\mu\nu}$ is the bosonic antisymmetric tensor
field (all with ghost number equal to zero), the scalar fields 
$\rho$ and  $\lambda$ are the fermionic (i.e. $\rho^2 = \lambda^2 = 0,
\rho \lambda + \lambda \rho = 0$) auxiliary (ghost) fields 
(with ghost number equal to $\mp 1$), the vector fermionic
fields  $(\bar C_\mu)C_\mu$ (i.e. $(C_\mu)^2 = (\bar C_\mu)^2 = 0,
C_\mu C_\nu + C_\nu C_\mu = 0, C_\mu \bar C_\nu + \bar C_\nu C_\mu = 0,
\bar C_\mu \bar C_\nu + \bar C_\nu \bar C_\mu = 0,$
etc.) are the (anti-)ghost fields (with ghost number $\mp 1$) and
bosonic fields $(\bar \beta)\beta$ are the bosonic (anti-)ghost fields
(with ghost number $\mp 2$). In fact, the above Lagrangian density
remains quasi-invariant
$$
\begin{array}{lcl}
s_{B} {\cal L}^{(l)}_{B} &=& \partial_\mu \;\bigl [\;
B^\mu \lambda - \rho \partial^\mu \beta + B_\nu (\partial^\mu C^\nu
- \partial^\mu C^\nu ) \;\bigr ] \nonumber\\
s_{D} {\cal L}^{(l)}_{B} &=& - \partial_\mu\; \bigl [\;
\lambda \partial^\mu \bar \beta + \rho {\cal B}^\mu +
{\cal B}_\nu (\partial^\mu \bar C^\nu - \partial^\nu \bar C^\mu)\; \bigr ]
\end{array}\eqno (2.2)
$$
under the following off-shell nilpotent ($s_{B}^2 = 0$) BRST ($s_{B}$)
$$
\begin{array}{lcl}
&&s_{B} B_{\mu\nu} = (\partial_{\mu} C_\nu - \partial_\nu C_\mu) \qquad
s_{B} C_\mu = \partial_\mu \beta \qquad s_{B} \beta = 0 \quad
s_{B} \bar C_\mu = B_\mu \nonumber\\ && s_{B} B_\mu = 0 \;
\qquad s_{B} \phi_{1} = - \lambda\; \qquad s_{B} \lambda = 0 \;\qquad
s_{B} \bar \beta = \rho \nonumber\\
&& s_{B} \rho = 0 \;\qquad s_{B} {\cal B}_\mu = 0 \qquad \;
s_{B} \phi_{2} = 0 \qquad \;s_{B} (\partial_{\mu} B_{\nu\kappa}) = 0
\end{array}\eqno(2.3a)
$$
and the off-shell nilpotent ($s_{D}^2 = 0$)
dual(co)-BRST ($s_{D}$) transformations
\footnote{ We adopt here the conventions and notations of [47] for the
(co-)BRST transformations. In fact, in its totality, a 
nilpotent ($\delta_{(D)B}^2 = 0$) (co-)BRST transformation
$\delta_{(D)B}$ is equal to the product of an anti-commuting 
number $\eta$
and the transformations $s_{(D)B}$ as: $\delta_{(D)B} = \eta s_{(D)B}$
with $s_{(D)B}^2 = 0$.}
$$
\begin{array}{lcl}
&&s_{D} B_{\mu\nu} =  \varepsilon_{\mu\nu\kappa\zeta} \partial^\kappa 
\bar C^\zeta  \qquad
s_{D} \bar C_\mu = - \partial_\mu \bar \beta \qquad 
s_{D} \bar \beta = 0 \qquad
s_{D} C_\mu = {\cal B}_\mu \nonumber\\
&& s_{D} {\cal B}_\mu = 0 \;\qquad
s_{D} \phi_{2} = - \rho \;\qquad s_{D} \rho = 0 \;\qquad
s_{D} \beta =  \lambda \nonumber\\
&& s_{D} \lambda = 0 \qquad\; s_{D} B_\mu = 0 \;\qquad
s_{D} \phi_{1} = 0 \qquad\; s_{D} (\partial^\mu B_{\mu\nu}) = 0.
\end{array}\eqno(2.3b)
$$

It can be readily checked that under the following discrete symmetry
transformations on the bosonic as well as fermionic fields of the theory
$$
\begin{array}{lcl}
&&\phi_{1} \rightarrow \pm i \phi_{2} \qquad 
\phi_{2} \rightarrow \mp i \phi_{1} \qquad 
B_{\mu} \rightarrow \pm i {\cal B}_\mu \nonumber\\
&&{\cal B}_{\mu} \rightarrow \mp i B_\mu \;\qquad
B_{\mu\nu} \rightarrow \mp \frac{i}{2} \varepsilon_{\mu\nu\kappa\zeta}
B^{\kappa\zeta}\nonumber\\ 
&&C_\mu \rightarrow \pm i \bar C_\mu 
\;\qquad \beta \rightarrow \mp i \bar \beta
\;\qquad \bar \beta \rightarrow \pm i  \beta  \nonumber\\
&& \rho \rightarrow  \pm i \lambda \;\qquad \lambda \rightarrow \pm i  \rho
\;\qquad \bar C_\mu \rightarrow \pm i C_\mu
\end{array}\eqno(2.4)
$$
the above Lagrangian density (2.1) remains invariant. At this juncture,
some of the salient points of our discussions,
are (i) the ghost part of the Lagrangian density
$ {\cal L}_{g} = - \partial_\mu \bar \beta \;\partial^\mu \beta
+ (\partial_\mu \bar C_\nu - \partial_\nu \bar C_\mu) (\partial^\mu C^\nu)
+ \rho \;(\partial_\mu C^\mu + \lambda) + (\partial_\mu \bar C^\mu + \rho)\;
\lambda $ remains invariant in itself under (2.4) but the kinetic energy term
and the gauge-fixing term exchange with each-other under 
the same transformations (i.e. eqn (2.4)). Exploiting
this observation, one can derive the off-shell nilpotent
$( s_{AB}^2 = s_{AD}^2 = 0$) anti-BRST ($s_{AB}$) and anti-co-BRST 
($s_{AD}$) symmetries from (2.3) by the substitution
$C_\mu \rightarrow \pm i \bar C_\mu, 
\bar C_\mu \rightarrow \pm i C_\mu,
\beta \rightarrow \mp i \bar \beta,
\bar \beta \rightarrow \pm i  \beta,  
\rho \rightarrow  \pm i \lambda, \lambda \rightarrow \pm i  \rho $. 
(ii) Under the (anti-)BRST and 
(anti-)co-BRST symmetry transformations, it is the kinetic energy term
(more precisely $ H = d B$) and the gauge-fixing term (i.e.
$ \partial^\kappa B_{\kappa\mu} dx^\mu = \delta B, \delta = \pm * d *$)
that are found to 
remain invariant, respectively. (iii) The anticommutator of the
(anti-)BRST and (anti-)co-BRST symmetries (i.e.
$\{s_{B}, s_{D} \} = \{s_{AB}, s_{AD} \} = s_{W}$) leads to the definition
of a bosonic symmetry transformation ($s_{W}$) as (see, e.g., [31-33])
$$
\begin{array}{lcl}
&& s_{W} B_{\mu\nu} = \partial_\mu {\cal B}_\nu - \partial_\nu {\cal B}_\mu
+ \varepsilon_{\mu\nu\kappa\xi} \partial^\kappa B^\xi \qquad
s_{W} C_\mu = \partial_\mu \lambda \nonumber\\
&& s_{W} (\phi_{1}, \phi_{2}, B_\mu, {\cal B}_\mu, \rho, \lambda, \beta,
\bar \beta ) = 0 \;\qquad \;
s_{W} \bar C_\mu = - \partial_\mu \rho \nonumber\\
&& s_{W} {\cal L}_{B} = \partial_\mu
[ \rho \partial^\mu \lambda - (\partial^\mu \rho)  \lambda + B^\kappa
\partial^\mu {\cal B}_\kappa - {\cal B}^\kappa \partial^\mu B_\kappa
+ {\cal B}^\mu (\partial_\kappa  B^\kappa) - B^\mu 
(\partial_\kappa {\cal B}^\kappa) ].
\end{array}\eqno(2.5)
$$
Under the above transformations,
 the ghost fields either do not transform or transform to
a vector gauge transformation (i.e. $s_{W} C_\mu = \partial_\mu \lambda,
s_{W} \bar C_\mu = - \partial_\mu \rho$). (iv) In addition to the above
transformations, there is a global {\it scale}
symmetry transformation (with global parameter $\Sigma$) which corresponds
to the ghost symmetry $s_{g}$. The infinitesimal version of this symmetry
is
$$
\begin{array}{lcl}
&& s_{g} B_{\mu\nu} = s_{g} B_\mu = s_{g} {\cal B}_\mu = s_{g} \phi_{1}
= s_{g} \phi_{2} = 0 \nonumber\\
&& s_{g} C_\mu = \Sigma C_\mu \qquad s_{g} \bar C_\mu = - \Sigma \bar C_\mu
\qquad  \;s_{g} \rho = - \Sigma \rho \nonumber\\
&& s_{g} \lambda  = \Sigma \lambda \;
\qquad s_{g} \bar \beta = - 2 \Sigma \bar \beta
\;\qquad  s_{g} \beta =  + 2  \Sigma \beta. 
\end{array}\eqno(2.6)
$$
Thus, there are six symmetries in the theory which are local, covariant 
as well as continuous and four (i.e. $s_{(A)B}^2 = 0, s_{(A)D}^2 = 0$)
of them are nilpotent of order two.\\

\noindent
{\bf 2.2 On-shell nilpotent symmetries}\\

\noindent
It is essential for our later discussions to elaborate on the presence of
on-shell nilpotent ($s_{(a)b}^2 = s_{(a)d}^2 = 0$) (anti-)BRST ($s_{(a)b}$)
and (anti-)co-BRST ($s_{(a)d}$) symmetries in the theory. For this purpose,
the following equations of motion, derived from (2.1)
$$
\begin{array}{lcl}
B_\mu &=& (\partial^\sigma B_{\sigma\mu} - \partial_\mu \phi_{1})\; \qquad
\;\lambda = - \frac{1}{2}\;(\partial \cdot C)
\equiv - \frac{1}{2} (\partial_\mu C^\mu) \nonumber\\
{\cal B}_\mu &=& (\frac{1}{2} \varepsilon_{\mu\nu\kappa\zeta}
\partial^\nu B^{\kappa\zeta} - \partial_\mu \phi_{2})\; \qquad
\rho = - \frac{1}{2}\;(\partial \cdot \bar C) \equiv - \frac{1}{2}
(\partial_\mu \bar C^\mu) 
\end{array}\eqno(2.7)
$$
play a very decisive role. In fact, it can be readily checked that,
with the help of (2.7), the Lagrangian density (2.1) can be recast as
$$
\begin{array}{lcl}
{\cal L}_{b}^{(0)} &=& 
 \frac{1}{2} (\partial^\sigma B_{\sigma\mu} - \partial_\mu
\phi_{1}) (\partial_\kappa B^{\kappa\mu} - \partial^\mu \phi_{1}) \nonumber\\
&-& \frac{1}{2} (\frac{1}{2} \varepsilon_{\mu\nu\kappa\zeta}
\partial^\nu B^{\kappa\zeta} - \partial_\mu \phi_{2}) 
(\frac{1}{2} \varepsilon^{\mu\xi\sigma\eta} \partial_\xi B_{\sigma\eta} -
\partial^\mu \phi_{2}) \nonumber\\
&-& \partial_\mu \bar \beta \partial^\mu \beta + (\partial_\mu \bar C_\nu
- \partial_\nu \bar C_\mu) (\partial^\mu C^\nu) - \frac{1}{2} (\partial \cdot
\bar C) (\partial \cdot C).
\end{array}\eqno(2.8)
$$
The equations of motion that emerge from the above Lagrangian density are
$$
\begin{array}{lcl}
\Box B_{\mu\nu} =
\Box \phi_{1} = \Box \phi_{2} =  \Box \beta = \Box \bar \beta = 0
 \quad \Box C_\mu = \frac{3}{2} \partial_\mu (\partial \cdot C)
\quad \Box \bar C_\mu = \frac{3}{2} \partial_\mu (\partial \cdot \bar C).
\end{array}\eqno(2.9)
$$
The Lagrangian density (2.8) is endowed with the following on-shell
nilpotent
($s_{b}^2 \phi_{1} \sim \Box \beta = 0, \;
s_{b}^2 \bar \beta \sim \Box \phi_{1} = 0, \;
s_{b}^2 \bar C_\mu \sim \Box C_\mu - \frac{3}{2} \partial_\mu 
(\partial \cdot C) = 0$) BRST symmetry transformations
$$
\begin{array}{lcl}
&&s_{b} B_{\mu\nu} = (\partial_{\mu} C_\nu - \partial_\nu C_\mu)\; \quad
s_{b} C_\mu = \partial_\mu \beta \; \quad s_{b} \beta = 0\; \quad
s_{b} \bar C_\mu = (\partial^\nu B_{\nu\mu} - \partial_\mu \phi_{1}) 
\nonumber\\ 
&&s_{b} \phi_{1} = \frac{1}{2} (\partial \cdot C)\; 
\qquad s_{b} \bar \beta = - \frac{1}{2} (\partial \cdot \bar C)\; 
\qquad s_{b} \phi_{2} = 0\; \qquad s_{b} (\partial_{\mu} B_{\nu\kappa}) = 0
\end{array}\eqno(2.10)
$$
together with the on-shell nilpotent
($ s_{d}^2 \phi_{2} \sim \Box \bar \beta = 0,  \;
s_{b}^2 \beta \sim \Box \phi_{2} = 0, \;
s_{b}^2 C_\mu \sim \Box \bar C_\mu - \frac{3}{2} \partial_\mu 
(\partial \cdot \bar C) = 0$) dual(co-)BRST transformations as listed below
$$
\begin{array}{lcl}
&&s_{d} B_{\mu\nu} =  \varepsilon_{\mu\nu\kappa\zeta} \partial^\kappa
\bar C^\zeta \; \qquad
s_{d} \bar C_\mu = - \partial_\mu \bar \beta\; 
\quad \;s_{d} C_\mu = (\frac{1}{2}
\varepsilon_{\mu\nu\kappa\sigma} \partial^\nu B^{\kappa\sigma}
- \partial_\mu \phi_{2}) \nonumber\\
&& s_{d} \bar \beta = 0\; \quad
s_{d} \phi_{2} = \frac{1}{2} (\partial \cdot \bar C)\; 
\quad s_{d} \beta = - \frac{1}{2} (\partial \cdot C)\; 
\quad s_{d} \phi_{1} = 0 \;\quad s_{d} (\partial^\kappa B_{\kappa\mu}) = 0.
\end{array}\eqno(2.11)
$$
Exploiting the discrete symmetry transformations of the ghost fields in
(2.4), it is straightforward to obtain the on-shell nilpotent
anti-BRST ($s_{ab}$) and anti-co-BRST 
($s_{ad}$) transformations from (2.10) and (2.11)
respectively. Similarly, the on-shell version of the bosonic symmetry ($s_{w}$)
transformations (2.5) can be obtained by substituting for the auxiliary
fields from equation (2.7). The ghost symmetry transformations
on the ghost fields in (2.6) remain intact (modulo the scale
transformations on the auxiliary fields
$B_\mu, {\cal B}_\mu, \rho, \lambda$ which no longer exist in the theory due
to the Euler-Lagrange equations of motion (2.7)). \\

\noindent
{\bf 3 Special features}\\

\noindent
We discuss here some of the distinguished properties of
the 4D free Abelian 2-form gauge theory keeping in mind similar
features in the context of one-form gauge theories.\\

\noindent
{\bf 3.1 Two-form gauge theory as the Hodge theory}\\

\noindent
For the present paper to be self-contained, we briefly mention some of the
key results of our earlier work [25] 
in a somewhat different manner than what has been discussed there. 
These results are intimately 
connected with the proof of the fact that the
4D 2-form gauge theory is a field theoretic model for the Hodge theory.
In this work [25], all the de Rham cohomology operators $(d,\delta,\Delta)$ are
identified with some specific symmetry transformations (and their generators)
for the Lagrangian density of the theory. It is straightforward to check
that all the six symmetries (on-shell as well as off-shell versions)
obey the following operator form of the algebra
\footnote{ We have taken here the notations of the on-shell version of the
symmetry transformations but the present algebraic structure is valid for
the off-shell version of the symmetry transformations as well. Furthermore,
a similar kind of operator algebra is obeyed by the off-shell as well as
on-shell version of the generators (cf (5.1) below) which are nothing but
the conserved Noether charges. It will be noted that (5.1), although
written for the off-shell version, is also true for
the on-shell version of the generators.}
$$
\begin{array}{lcl}
&& s_{(a)b}^2 = 0 \qquad s_{(a)d}^2 = 0 \qquad
[ s_{w}, s_{r} ] = 0 \qquad r = b, ab, d, ad, g \nonumber\\
&& \{ s_{b}, s_{ab} \} = 0 \;\qquad 
 \{ s_{d}, s_{ad} \} = 0\; \qquad 
i [ s_{g}, s_{b(ad)} ] = + s_{b(ad)}
\nonumber\\ 
&& i [ s_{g}, s_{d(ab)} ] = - s_{b(ad)} \;\qquad
\{ s_{b}, s_{d} \} = s_{w} = \{ s_{ab}, s_{ad} \} 
\end{array}\eqno(3.1)
$$
which is reminiscent of the algebraic structure obeyed by the
cohomological operators $(d,\delta, \Delta)$. A noteworthy point is
the fact that the discrete symmetry transformations (2.4)
correspond to the Hodge duality $*$ operation of the differential
geometry on a compact manifold. In this 
context, it can be checked that the following 
relationships between the (anti-)co-BRST and (anti-)BRST transformations
are satisfied:
$$
\begin{array}{lcl}
s_{(d, D)} \; \Phi = \pm \; *\; s_{(b, B)}\; *\; \Phi\; \;\qquad\;\;
s_{(ad, AD)} \;\Phi = \pm \; *\; s_{(ab, AB)}\; *\; \Phi 
\end{array}\eqno(3.2)
$$
where $\Phi$ stands for the generic field of the Lagrangian
densities (2.1) and (2.8). In the above, the $\pm$ signs for a 
particular field are dictated by the corresponding signs that appear
when we take two successive $*$ operation on that field [48-50]. 
In more general terms,
this statement can be expressed
succinctly, for the generic field $\Phi$ of the theory, as
$$
\begin{array}{lcl}
*\; (\; * \; \Phi) = \pm\; \Phi
\end{array}\eqno(3.3)
$$
where, as is evident, the $*$ operation is nothing but the discrete
symmetry transformations in (2.4) for the fields. For the theory under 
consideration, it turns out that
$$
\begin{array}{lcl}
*\; ( *\; B ) &=& + \; B\; \qquad B = \beta, \bar \beta, \phi_{1},
\phi_{2}, B_{\mu\nu}, B_\mu, {\cal B}_\mu \nonumber\\
 {*}\; ( {*}\; F ) &=& - \; F\; 
\qquad F = \rho, \lambda, C_\mu, \bar C_\mu.
\end{array}\eqno(3.4)
$$
Thus, the $(\pm)$ signs in (3.2) find their physical interpretation in the
sense that the nilpotent (anti-)co-BRST and (anti-)BRST transformations on the
bosonic(fermionic) fields $(B)F$ of the 2-form gauge theory 
are related through the interplay between discrete and continuous
symmetries of the theory. In explicit form, the above
relationship can be expressed as:
$ s_{(d, D)} B = + * s_{(b, B)} * B,
 s_{(d, D)} F = - * s_{(b, B)} * F$ where $B$ and $F$ of the theory
are defined in (3.4). As is self evident, a similar kind of relationship
exists for the anti-BRST and anti-co-BRST transformations if we could
exploit (2.4) judiciously.

The relationship in (3.2) is exactly same as
the connection between cohomological operators $\delta$ and $d$
(i.e. $ \delta = \pm \;*\; d\; *$). Whereas in differential geometry, the 
$(\pm)$ signs in $\delta = \pm * d *$ are dictated by the dimensionality
\footnote{In any arbitrary {\it even} $D$-dimensional manifold, $d$ and
$\delta$ are related by: $ \delta = - * d *$. In general, an inner product
of a n-form in $D$ dimensional manifold leads to the relationship between $d$
and $\delta$ as:
$\delta = (-1)^{Dn + D + 1} * d *$. Thus, for the {\it odd } dimensional
manifold, we have $\delta = (-1)^{n} * d *$ (see, e.g., [26,27]).}
of the manifold on which the Hodge duality is defined (see, e.g., [26-30]),
it is evident that, in the context of 2-form gauge theory these signs are
dictated by (3.3). Ultimately, it so happens that
the signs in (3.3) are
guided by the nature of field(s) of the theory under consideration.
Other useful properties of the 2-form Abelian gauge theory, under the
$*$ operation (corresponding to the discrete symmetry transformation
(2.4)), are: (i) the Lagrangian density (2.1) remains invariant. The
on-shell version of the (2.4) can be obtained by substituting for
the auxiliary fields of the theory from (2.7). Under the latter
transformations, the
Lagrangian density (2.8) remains invariant. (ii) The continuous symmetry
transformations (on-shell as well as off-shell versions)
and corresponding symmetry generators
$Q_{r}, r = B (b), AB(ab), D(d), AD(ad), g$ of the theory 
transform, under (2.4), as
$$
\begin{array}{lcl}
&&s_{(A)B} \rightarrow s_{(A)D}\; \quad
s_{(A)D} \rightarrow - s_{(A)B}\; \quad
s_{(W,w)} \rightarrow - s_{(W,w)}\; \quad s_{g} \rightarrow - s_{g} \nonumber\\
&& Q_{(A)B} \rightarrow Q_{(A)D}\; \quad
Q_{(A)D} \rightarrow - Q_{(A)B}\; \quad
Q_{(W,w)} \rightarrow - Q_{(W,w)}\; \quad Q_{g} \rightarrow - Q_{g} \nonumber\\
&&s_{(a)b} \rightarrow s_{(a)d}\; \quad
s_{(a)d} \rightarrow - s_{(a)b}\; \quad
 Q_{(a)b} \rightarrow Q_{(a)d}\; \quad
Q_{(a)d} \rightarrow - Q_{(a)b} 
\end{array}\eqno(3.5)
$$
where the local expressions for the conserved charges $Q_{r}$ can be found
in [25]. For our present discussions, these expressions are not required
to be quoted here.
(iii) The above transformations are such that the algebraic structure of
(3.1) does not change at all. (iv) There exists an inverse relationship
{\it vis-{\`a}-vis} relations (3.2) as given below
$$
\begin{array}{lcl}
s_{(b, B)} \; \Phi = \mp \; *\; s_{(d, D)}\; *\; \Phi \; \qquad \;
s_{(ab, AB)} \;\Phi = \mp \; *\; s_{(ad, AD)}\; *\; \Phi 
\end{array}\eqno(3.6)
$$
where, it can be noted, the signs have changed in the r.h.s. In explicit
form, the above transformations can be expressed as:
$ s_{(b,B)} B  = - * s_{(d, D)} * B, s_{(b,B)} F = + * s_{(d,D)} * F$
where $B$ and $F$ are the bosonic and fermionic fields of the theory. 
This fact should
be contrasted with the case of 2D free Abelian gauge theory, for which, the
signs are same in the equations corresponding to
(3.2) and (3.6) [33]. This peculiarity happens
because the duality in 4D and 2D are different [48-50]. (v) The
transformations $s_{(b,B)} \rightarrow s_{(d, D)}, \; s_{(d, D)} \rightarrow
- s_{(b, B)}$ under the Hodge $*$ operation are reminiscent of the duality
transformations for the electric (${\bf E}$)
and magnetic (${\bf B}$) fields in the case of source
free Maxwell equations where $ {\bf E} \rightarrow {\bf B}, {\bf B}
\rightarrow - {\bf E}$.

All the above properties establish the fact that the 4D free Abelian
2-form gauge theory is a field theoretical model for the Hodge theory
because the local, covariant and continuous symmetry transformations 
(the off-shell as well as the on-shell versions) for the
Lagrangian density of the theory 
(and the corresponding Noether conserved charges) 
obey an algebra that is reminiscent
of the algebra obeyed by the de Rham cohomological operators 
$(d, \delta,\Delta)$. The discrete symmetry transformations (2.4)
(and its on-shell version) are the analogue of the Hodge duality $*$
operation of the differential geometry. As far as the ghost number
consideration of the states in the quantum Hilbert space are concerned,
it turns out that the set $(d,\delta,\Delta)$ has a one-to-two mappings
with the conserved charges (or generators) of the theory, namely;
$d \rightarrow Q_{b(ad)}, \delta \rightarrow Q_{d(ab)}, \Delta
\rightarrow Q_{w} = \{ Q_{b}, Q_{d} \} = \{ Q_{ab}, Q_{ad} \}$. The basic
theoretical reason behind the existence of such a mapping
can be explained in the framework of
superfield formulation of the present theory as has been done for the
one-form free Abelian (and self-interacting non-Abelian) gauge theory
(see, e.g., [38-43] for details). Because of the existence of the above mapping,
the analogue of the Hodge decomposition theorem can be defined in the
quantum Hilbert space of states [25]. In fact, all the above cited properties
of this section are common to the 2D one-form free Abelian (and 
self-interacting non-Abelian) as well as 4D free Abelian
2-form gauge theories [25,31-36] except for the sign difference in the 
inverse relationship (3.6). 
To be more precise, the analogues of (3.2) and (3.6) for the 2D theories
bear the same signs on the r.h.s. in contrast to the 4D theories where
there is a sign flip on the r.h.s. of the relationships (3.2) and (3.6).\\

\noindent
{\bf 3.2 Towards topological aspects}\\

\noindent
It can be checked explicitly that the linearized version of the
the Lagrangian density (2.1) can be expressed in the
following two different ways by exploiting the off-shell nilpotent
($s_{B}^2 = s_{D}^2 = 0$) (co-)BRST symmetries ($s_{(D)B}$)
(cf (2.3)) of the theory
$$
\begin{array}{lcl}
{\cal L}^{(l)}_{B} &=& 
 B^\mu (\partial^\sigma B_{\sigma \mu} - \partial_\mu \phi_{1})
- \frac{1}{2}\; B^\mu B_\mu  + s_{D}\; (G) + \partial_\mu \;Z^\mu
\end{array}\eqno(3.7a)
$$
$$
\begin{array}{lcl}
{\cal L}^{(l)}_{B} &=& 
 \frac{1}{2}\; {\cal B}^\mu {\cal B}_\mu 
- {\cal B}^\mu (\frac{1}{2}\; 
\varepsilon_{\mu\nu\kappa\zeta} \partial^\nu B^{\kappa\zeta} 
- \partial_\mu \phi_{2})
+ s_{B}\; (F) + \partial_\mu\; T^\mu
\end{array}\eqno(3.7b)
$$
where the expressions for $G, Z^\mu, F, T^\mu,$ in terms of the local fields, 
are
$$
\begin{array}{lcl}
G &=& \frac{1}{2} C^\mu ({\cal B}_\mu - \varepsilon_{\mu\nu\kappa\zeta}
\partial^\nu B^{\kappa\zeta}) - (\partial \cdot C + \lambda) \phi_{2}
- (\partial \cdot \bar C + \rho) \beta \nonumber\\
Z^\mu &=& {\cal B}^\mu \phi_{2} - \beta \partial^\mu \bar\beta +
(\partial^\mu \bar C^\nu - \partial^\nu \bar C^\mu ) C_\nu \nonumber\\
F &=& \bar C^\mu (\partial^\sigma B_{\sigma\mu} - \frac{1}{2} B_\mu)
+ (\partial \cdot C + \lambda) \bar \beta
+ (\partial \cdot \bar C + \rho) \phi_{1} \nonumber\\
T^\mu &=& \bar C_\nu (\partial^\mu C^\nu - \partial^\nu C^\mu)
- B^\mu \phi_{1} - \bar\beta \partial^\mu \beta.
\end{array}\eqno(3.8)
$$
The above equation (3.7) can also be expressed in terms of the anti-BRST
and anti-co-BRST transformations by exploiting the discrete
transformations (2.4) for the ghost fields. 
The analogue of expressions in (3.8),
for the latter case, can also be computed by the substitution of
transformations for the ghost fields in (2.4). Thus, it is evident that
there are four ways to express the Lagrangian density (2.1) in terms of
the four off-shell nilpotent symmetry transformations ($s_{(A)B}, s_{(A)D}$)
of the theory. In fact, this special feature of the 2-form gauge theory
is identical to the 2D free Abelian (and self-interacting non-Abelian) 1-form
gauge theories which have been shown [33,39-41] to be topological in nature.

One of the interesting features of a field theory to be a
topological field theory is the existence of topological invariants which
obey a certain specific type of recursions relations (see, e.g.,
[37] for details). In fact, it has
been shown for the one-form free 2D Abelian gauge theory
(as well as the self-interacting 2D non-Abelian gauge theory) that there are
four sets of such invariants $(\tilde I_{k})I_{k} $ and $(\tilde J_{k}) J_{k}$
w.r.t. the four conserved ($\dot Q_{(A)B} = \dot Q_{(A)D} = 0$)
and nilpotent ($ Q_{(A)B}^2 = Q_{(A)D}^2 = 0$) (anti-)BRST
charges $Q_{(A)B}$ and (anti-)co-BRST charges $Q_{(A)D}$, namely;  
$$
\begin{array}{lcl}
I_{k} = {\displaystyle \oint}_{C_{k}} V_{k} \;\qquad
\tilde I_{k} = {\displaystyle \oint}_{C_{k}} \tilde V_{k} \;\qquad
J_{k} = {\displaystyle \oint}_{C_{k}}  W_{k} \;\qquad
\tilde J_{k} = {\displaystyle \oint}_{C_{k}} \tilde W_{k} 
\end{array}\eqno(3.9)
$$
where $(\tilde V_{k})V_{k}$ and $(\tilde W_{k}) W_{k}$
are the k-forms ($ k = 0, 1, 2)$ and $C_{k}$ are the homology cycles on
the 2D closed Riemann surface which is an Euclidean version of the 2D Minkowski
manifold we started with. Euclideanization is essential to have meaning
of the topological invariants, homology cycles etc. in the language of the
curves in algebraic geometry as the Minkowski manifold is
a non-compact manifold [51]. Since there is a single
set of (anti-)ghost fields $(\bar C)C$ in one-form 2D (non-)Abelian
gauge theories, we
obtain first the two sets of invariants w.r.t. (co-)BRST charges. From these,
the other two  (w.r.t. anti-BRST and anti-co-BRST charges)
can be computed by exploiting the discrete ghost symmetries of the 
theory [31-34]. In contrast, in the present
2-form 4D Abelian gauge theory, there are three sets of (anti-)ghost
fields, namely; $(\bar\beta) \beta, (\bar C_\mu) C_\mu, (\rho) \lambda$
with ghost numbers: $(-2)2, (-1)1, (-1)1$, respectively. 
Thus we, expect the existence of invariants including all
these ghost fields. It can be checked that the following set
(which starts with an invariant containing $\beta$)
$$
\begin{array}{lcl}
V_{0} (+2) &=& (\partial \cdot B) \beta\; \nonumber\\
V_{1} (+1) &=& \bigl [ \partial_\mu (\partial \cdot \bar C) \beta
+ (\partial \cdot B)\; C_\mu \bigr ]\; dx^\mu\; \nonumber\\
V_{2} (0) &=& \bigl [ \partial_\mu (\partial \cdot \bar C) C_\nu
+ \frac{1}{2} (\partial \cdot B) B_{\mu\nu} \bigr ]\;
dx^\mu \wedge dx^\nu\; \nonumber\\
V_{3} (-1) &=& \bigl [ \frac{1}{2} \partial_\mu (\partial \cdot \bar C)
B_{\nu\lambda} + \frac{1}{3!} (\partial \cdot \bar C) H_{\mu\nu\kappa}
\bigr ]\; dx^\mu \wedge dx^\nu \wedge dx^\kappa \nonumber\\
&\equiv& d\; \bigl [ \frac{1}{2} (\partial \cdot \bar C) B_{\nu\kappa}
\; dx^\nu \wedge dx^\kappa \bigr ]\; \nonumber\\
V_{4}  &=& 0\; \qquad ( d^2 = 0 ).
\end{array}\eqno(3.10)
$$
is a set of invariants w.r.t. the conserved
and nilpotent BRST charge $Q_{B}$. The noteworthy points, at this stage,
are: (i) the zero-form $V_{0}$ is a BRST invariant quantity
(i.e. $s_{B} V_{0} = 0$). (ii) The three form $V_{3}$ turns out to be
an exact form. Thus, we are unable to deduce the four-form $V_{4}$ for 
the theory on 4D manifold. (iii) The ghost numbers for the 
forms $V_{0}-V_{3}$ are $ + 2, +1, 0, -1$ respectively. These
numbers have been exploited to express uniquely the $k$-forms
$V_{k} (n), k = 0, 1, 2, 3$ for $n = +2, +1, 0, -1$. (iv) The above forms
obey a recursion relation (a key characteristic feature of an exact TFT) as 
$$
\begin{array}{lcl}
s_{B} I_{k} = d \; I_{k - 1}\; \qquad (k = 1, 2, 3, 4)
\end{array}\eqno(3.11)
$$
where $d = dx^\mu \partial_\mu$ is the exterior derivative. (v) From the
above k-forms $V_{k}$, one can compute easily the $k$-forms $\tilde V_{k}$
corresponding to the nilpotent anti-BRST charge $Q_{AB}$ by exploiting the
discrete transformations for the ghost fields:
$\beta \rightarrow \mp i \bar \beta, \bar \beta \rightarrow \pm i \beta,
\rho \rightarrow \pm i \lambda, \lambda \rightarrow \pm i \rho,
C_\mu \rightarrow \pm i \bar C_\mu, \bar C_\mu \rightarrow \pm i C_\mu$.
The resulting invariants $\tilde I_{k}, (k = 0, 1, 2, 3, 4)$ obey the
same kind of recursion relations as (3.11) (i.e.
$s_{AB} \tilde I_{k} = d \; \tilde I_{k - 1}, (k = 1, 2, 3, 4))$.

One can now compute the invariants w.r.t. the conserved and nilpotent
co-BRST charge $Q_{D}$. The set $W_{k} (k = 0,1,2,3,4)$,
which constitutes the invariants $J_{k}$, is as follows
$$
\begin{array}{lcl}
W_{0} (-2) &=& (\partial \cdot {\cal B}) \bar \beta \nonumber\\
W_{1} (-1) &=& \bigl [ \partial_\mu (\partial \cdot C) \bar \beta
- (\partial \cdot {\cal B})\; \bar C_\mu \bigr ]\; dx^\mu \nonumber\\
W_{2} (0) &=& \bigl [ - \partial_\mu (\partial \cdot C) \bar C_\nu
+ \frac{1}{4} (\partial \cdot {\cal B}) \varepsilon_{\mu\nu\kappa\sigma}
B^{\kappa\sigma}  \bigr ]\;
dx^\mu \wedge dx^\nu \nonumber\\
W_{3} (+1) &=& \frac{1}{4} \bigl [ \partial_\mu (\partial \cdot C)
B^{\kappa\sigma} + (\partial \cdot \bar C) \partial_\mu B^{\kappa\sigma}
\bigr ]\; \varepsilon_{\nu\eta\kappa\sigma}
dx^\mu \wedge dx^\nu \wedge dx^\eta \nonumber\\
&\equiv& d\; \bigl [ \frac{1}{4} (\partial \cdot \bar C) 
\varepsilon_{\mu\nu\kappa\sigma} B^{\kappa\sigma}
\; dx^\mu \wedge dx^\nu \bigr ] \nonumber\\
W_{4}  &=& 0\; \qquad ( d^2 = 0 ).
\end{array}\eqno(3.12)
$$
Some of the salient features, at this juncture, are (i) from the above forms,
one can compute the forms $\tilde W_{k} (k = 0, 1, 2, 3, 4) $ of (3.9) by 
exploiting the discrete transformations on the bosonic as well as fermionic
ghost fields of equation (2.4). (ii) The zero-forms $\tilde W_{0} (+2)$
and $W_{0} (-2)$ are invariant under the (anti-)co-BRST transformations
$s_{(A)D}$ because $s_{AD} \tilde W_{0} (+2) = 0, s_{D} W_{0} (-2)
= 0$. (iii) The ghost numbers for $J_{0}-J_{3}$,
constructed from $W_{0}-W_{3}$, are $-2, -1, 0, +1$
respectively. (iv) It is interesting to note that
$$
\begin{array}{lcl}
*\; V_{k} (n) = W_{k} (\tilde n)\; \qquad
*\; \tilde V_{k} (\tilde n) = \tilde W_{k} (n)\; \qquad
\tilde n = -2, \mp 1, 0\; \qquad n = +2, \pm 1, 0
\end{array}\eqno(3.13)
$$
where the $*$ operation corresponds to the discrete symmetry transformations 
in (2.4). (v) The invariants $J_{k}, \tilde J_{k}$, constructed from
$ W_{k} (\tilde n), \tilde W_{k} (n)$, obey the following recursion 
relations
$$
\begin{array}{lcl}
s_{D} J_{k} = d \; J_{k - 1}\; \qquad 
s_{AD} \tilde J_{k} = d \; \tilde J_{k-1}\; \qquad (k = 1, 2, 3, 4)
\end{array}\eqno(3.14)
$$
which is the analogue of the similar relation in (3.11) w.r.t. the BRST
charge $Q_{B}$.

Taking into account now the bosonic (anti-)ghost fields $(\bar\beta)\beta$
and the fermionic vector (anti-)ghost fields $(\bar C_\mu)C_\mu$, we can
construct the following zero-forms $W_{0}$ and $V_{0}$ 
$$
\begin{array}{lcl}
W_{0} (-1) = - [ {\cal B}_\mu \bar C^\mu + C^\mu \partial_\mu \bar \beta ]\;
\;\;\qquad \;\;V_{0} (+1) = B_{\mu} C^\mu - \bar C^\mu \partial_\mu \beta
\end{array}\eqno(3.15)
$$
that remain invariant under the (co-)BRST  transformations $s_{(D)B}$ 
as $s_{D} W_{0} = 0, s_{B} V_{0} = 0$. In the above equation, the numbers
in the round bracket (i.e. $(\pm 1)$) denote the ghost numbers for the
zero-forms. The other non-trivial forms $V_{k}, k = 1, 2$, that are found to
be consistent with the recursion relations (3.11) and can be constructed
(with ghost numbers $0, -1$) in terms of the (anti-)ghost fields 
$(\bar C_\mu)C_\mu$, are                    
$$
\begin{array}{lcl}
V_{1} (0) &=& \bigl [ (\partial_\kappa \bar C_\mu) C^\mu + \bar C^\mu
\partial_\mu C_{\kappa} + B^\mu B_{\kappa\mu} \bigr ] \; dx^\kappa \nonumber\\
V_{2} (-1) &=& \bigl [ (\partial_\sigma \bar C^\mu) B_{\kappa\mu}
+ \bar C^\mu (\partial_\sigma B_{\kappa\mu}) \bigr ]\; dx^\sigma \wedge
dx^\kappa \nonumber\\
&\equiv& d\; \bigl [ (\bar C^\mu B_{\kappa\mu}) dx^\kappa \bigr ].
\end{array}\eqno(3.16)
$$
It is evident that $V_{3} = V_{4} = 0$ as $V_{2} (-1)$ turns out to be
an exact form. Thus, recursion relations (cf (3.11))
allow the existence of the forms
only up to {\it degree two}
 because, at order two itself, we obtain an exact form.
It should be noted, at this stage, that there is another candidate 
$V_{1}^{(a)} (0)$ for degree one, namely;
$$
\begin{array}{lcl}
V_{1}^{(a)} (0) = \bigl [\; (\partial_\kappa \bar C_\mu) C^\mu
+ \bar C^\mu (\partial_\kappa C_\mu) \;\bigr ] \;dx^\kappa \equiv
d\; [\;\bar C_\mu C^\mu \;]
\end{array}\eqno(3.17)
$$
which obeys the recursion relations (3.11). But we do not consider it because
it is already an exact form (and, therefore, trivial). Now, 
consistent with the recursion relations (3.14), the higher order forms
with respect to the co-BRST charge $Q_{D}$ are
$$
\begin{array}{lcl}
W_{1} (0) &=& - \bigl [ (\partial_\kappa C_\mu) \bar C^\mu
+ C^\mu (\partial_\mu \bar C_\kappa) - \frac{1}{2} 
\varepsilon_{\kappa\mu\eta\zeta} {\cal B}^\mu B^{\eta\zeta} \bigr ]\; 
dx^\kappa \nonumber\\
W_{2} (-1) &=& \frac{1}{2}\; \bigl [ (\partial_\sigma C^\mu) B^{\eta\zeta}
+ C^\mu \partial_\sigma B^{\eta\zeta} \bigr ]\; 
\varepsilon_{\kappa\mu\eta\zeta} \; dx^\sigma \wedge dx^\kappa
\nonumber\\
&\equiv & \frac{1}{2} d\; \bigl [ \varepsilon_{\kappa\mu\eta\zeta} 
C^\mu B^{\eta\zeta}  dx^\kappa \bigr ].
\end{array}\eqno(3.18)
$$
It is very much clear that $ W_{3} = W_{4} = 0$ because the 2-form
$W_{2} (-1)$ is itself an exact form. Thus,
the series of invariants terminates here and $W_{3}$ and $W_{4}$ turn
out to be trivially zero (because $d^2 = 0$). Analogous to (3.17), there
is an additional candidate $W_{1}^{(a)} (0)$ 
but it turns out to be an exact form (i.e. $ W_{1}^{(a)} 
= - d [C_\mu \bar C^\mu])$. This is why, it has {\it not}
 been considered in the
above series. It is obvious that the forms $\tilde V_{k}, \tilde W_{k}
(k = 0, 1, 2)$ with respect to the anti-BRST ($Q_{AB}$)
and anti-co-BRST ($Q_{AD})$ charges
can be computed from (3.15)--(3.18) by exploiting the discrete symmetry
transformations on the ghost fields in (2.4). The ghost numbers for the
invariants constructed from $\tilde V_{k}$ and $\tilde W_{k}$ would
turn out to be $(-1, 0, +1)$ and $(+1, 0, -1)$ respectively and they would
obey the recursion relations $s_{AB} \tilde I_{k} = d I_{k-1},
s_{AD} \tilde J_{k} = d \tilde J_{k-1}, (k = 1, 2, 3,)$.

Let us now concentrate on the invariants that are constructed with the 
ghost field $\lambda$ which carries the ghost number equal to $+1$. The
following set
$$
\begin{array}{lcl}
V_{0} (+1) &=& (\partial \cdot B) \lambda \nonumber\\
V_{1} (0) &=& \bigl [ \partial_\mu (\partial \cdot \bar C) \lambda
+ (\partial \cdot B)\; \partial_\mu \phi_{1}\bigr ]\; dx^\mu \nonumber\\
V_{2} (-1) &=& - \bigl [ \partial_\mu (\partial \cdot \bar C) \partial_\nu
\phi_{1} \;\bigr ]\;
dx^\mu \wedge dx^\nu 
\equiv - d\; \bigl [ (\partial \cdot \bar C) \partial_{\nu} \phi_{1}
\; dx^\nu \bigr ] \nonumber\\
V_{3} = V_{4}  &=& 0 \;\qquad \; ( d^2 = 0 )
\end{array}\eqno(3.19)
$$
obeys the recursion relations given in (3.11). From the BRST invariant 
($s_{B} V_{0} (+1) = 0$) zero-form $V_{0}(+1)$
in the above, one can construct an 
additional set $V^{(a)}_{k} (k = 1, 2)$ as 
$$
\begin{array}{lcl}
V^{(a)}_{1} (0) &=& \bigl [ (\partial \cdot \bar C) \partial_\mu \lambda
+ \partial_\mu (\partial \cdot B)\; \phi_{1}\bigr ]\; dx^\mu \nonumber\\
V^{(a)}_{2} (-1) &=&  \bigl [ \partial_\mu (\partial \cdot \bar C) \partial_\nu
\phi_{1} \;\bigr ]\;
dx^\mu \wedge dx^\nu
\equiv + d\; \bigl [ (\partial \cdot \bar C) \partial_{\nu} \phi_{1}
\; dx^\nu \bigr ] \nonumber\\
V_{3}^{(a)} = V_{4}^{(a)}  &=& 0\; \qquad (d^2 = 0).
\end{array}\eqno(3.20)
$$
The above set also obeys the recursion
relations of (3.11). It should be emphasized that there are two more candidates
for the one-form $V_{1} (0)$ (with ghost number $0$) which obey the
recursion relations (3.11). But, they turn out to be trivial in the sense that
they are found to be exact forms. The precise expressions for them are:
$V^{(1)}_{1} (0) = d [(\partial \cdot \bar C) \lambda],
V^{(2)}_{1} (0) = - d [(\partial \cdot B) \phi_{1}]$. It is evident that,
w.r.t. these trivial one-forms, there are no non-trivial 2-forms in
the recursion relations (3.11) because $(d^2 = 0)$. Analogous to the above,
a set of invariants can be constructed w.r.t. conserved 
$(\dot Q_{D} = 0)$ and nilpotent $(Q_{D}^2 = 0)$
co-BRST charge $Q_{D}$. The corresponding forms
$W_{k}, (k = 0, 1, 2, 3, 4)$, with self-evident notations,  are 
$$
\begin{array}{lcl}
W_{0} (-1) &=& (\partial \cdot {\cal B}) \rho \nonumber\\
W_{1} (0) &=& - \bigl [ \partial_\mu (\partial \cdot C) \rho
- (\partial \cdot {\cal B})\; \partial_\mu \phi_{2}\bigr ]\; dx^\mu
\nonumber\\
W_{2} (+1) &=& \bigl [ \partial_\mu (\partial \cdot C) \partial_\nu \phi_{2}
\bigr ]\;
dx^\mu \wedge dx^\nu
\equiv d\; \bigl [ (\partial \cdot C) 
\partial_{\nu} \phi_{2} dx^\nu \bigr ] \nonumber\\
W_{3} = W_{4}  &=& 0\; \qquad \;( d^2 = 0 ).
\end{array}\eqno(3.21)
$$
The invariants $J_{k}$  constructed from the above forms
obey the recursion relations given in (3.14). There exist additional
forms $W^{(a)}_{k}$ which can be constructed from the co-BRST invariant
($s_{D} W_{0} (-1) = 0$) zero-form $W_{0}(-1)$. These are as follows
$$
\begin{array}{lcl}
W^{(a)}_{1} (0) &=& - \bigl [ (\partial \cdot C) \partial_\mu \rho
- \partial_\mu (\partial \cdot {\cal B})\; \phi_{2}\bigr ]\; dx^\mu
\nonumber\\
W^{(a)}_{2} (+1) &=& -
\bigl [ \partial_\mu (\partial \cdot C) \partial_\nu \phi_{2}
\bigr ]\; dx^\mu \wedge dx^\nu
\equiv - d\; \bigl [ (\partial \cdot C) 
\partial_{\nu} \phi_{2} dx^\nu \bigr ] \nonumber\\
W^{(a)}_{3} = W^{(a)}_{4}  &=& 0\; \qquad \;( d^2 = 0 ).
\end{array}\eqno(3.22)                                                          
$$
The invariants constructed from the above also obey (3.14). There are a few
comments in order now. (i) It is straightforward to check that the forms
$\tilde W_{k} (k = 0, 1, 2, 3, 4)$ corresponding to the conserved 
($\dot Q_{AD} = 0$) and nilpotent ($Q_{AD}^2 = 0$) anti-dual BRST
charge $Q_{AD}$ can be derived from $W_{k}$ and $W_{k}^{(a)}$ by exploiting
the discrete symmetry transformations on the ghost fields in (2.4).
(ii) The invariants $\tilde J_{k} (k = 0, 1, 2, 3, ,4)$ constructed from
these forms would obey the same kind of recursion relations as given in (3.14).
(iii) It is interesting to note that if  we exploit the full discrete
transformations of the analogue of Hodge duality $*$ operation in (2.4), we
obtain
$$
\begin{array}{lcl}
* \; V_{k} (n) &=& W_{k} (\tilde n)\; \qquad 
*\; \tilde V_{k} (\tilde n) = \tilde W_{k} (n) \nonumber\\
{*} \; V^{(a)}_{k} (n) &=& W^{(a)}_{k} (\tilde n)\; \qquad 
*\; \tilde V^{(a)}_{k} (\tilde n) = \tilde W^{(a)}_{k} (n) 
\end{array}\eqno(3.23)
$$
where $ n = +1, 0, -1$ and $\tilde n = -1, 0, +1$.\\

\noindent
{\bf 4 One-form 2D- versus two-form 4D free Abelian gauge theories}\\

\noindent
In this section, we shall compare and contrast between some of the key issues
associated with 2D free vector Abelian gauge theory {\it vis-{\`a}-vis} 4D
free Abelian antisymmetric gauge theory. For the sake of completeness
and clarity, we first recapitulate some of the key properties of the
2D one-form Abelian gauge theory  
\footnote{ We shall not mention, in this section, similar 
properties associated with the 2D self-interacting non-Abelian
gauge theory. However, these features can be seen in detail in
[33,34,40,42].} [31-33]. 
Before we embark on addressing the difference(s)
between these theories, it is important to emphasize that these theories
have something in common. These key similarities are: (i) they are 
field theoretical models for the Hodge theory because the symmetries 
(and corresponding generators) of the Lagrangian density obey the same
kind of algebra as obeyed by the de Rham cohomology operators 
$(d, \delta, \Delta)$.
(ii) The analogue of the Hodge duality $*$ transformations in (2.4) can 
be defined for the one-form gauge theory as well [33].
(iii) The (anti-)BRST and (anti-)co-BRST invariant quantities obey the same
kind of recursion relations as that of a typical field theoretical examples
of an exact TFT. (iv) These invariants terminate after reaching the 
differential-form 
that could be defined on the specific manifold with the 
maximum degree. Now, we concentrate on
the discussion of a key difference between these theories. To this end in
mind, we begin with the BRST invariant Lagrangian
density for the free 2D Abelian gauge theory in the Feynman gauge
(see, e.g. [20,47,52,53])
$$
\begin{array}{lcl}
\tilde {\cal L}_{b} = - \frac{1}{4} F_{\mu\nu} F^{\mu\nu} - \frac{1}{2}
(\partial \cdot A)^2 - i \partial_\mu \bar C \partial^\mu C
\equiv
 \frac{1}{2} E^2 - \frac{1}{2} 
(\partial \cdot A)^2 - i \partial_\mu \bar C \partial^\mu C
\end{array}\eqno(4.1)
$$
where only the electric field component $F_{01} = - \varepsilon^{\mu\nu}
\partial_\mu A_\nu = E$ (with $\varepsilon_{01} = \varepsilon^{10} = + 1$)
of the curvature tensor $F_{\mu\nu}$ exists in 2D
and $F_{\mu\nu}$ is derived from the 2-form $F = d A$
(with $d = dx^\mu \partial_\mu, d^2 = 0$ and $ A = dx^\mu A_\mu$).
Here the gauge-fixing term $(\partial \cdot A) = \delta A 
(\delta = - * d *, \delta^2 = 0)$ 
is derived from the co-exterior derivative
\footnote{ The one-form $A = A_{\mu} dx^\mu$ defines the vector
potential for the Abelian gauge theory. The zero-form (gauge-fixing) 
$\delta A = (\partial \cdot A)$ and the curvature 2-form 
(field strength tensor) $F^{(A)} = d A$ are 
`Hodge dual' to each-other in any arbitrary dimension of spacetime. Here 
$\delta = \pm * d *$ is the co-exterior derivative w.r.t. $d$. The same
is not true ($ F^{(N)} \neq d A$) for the non-Abelian gauge theory
(see, e.g. [26-30]).}  
$\delta$ which is the `Hodge dual' of $d$ and (anti-)ghost fields $(\bar C)C$
are anti-commuting ($\bar C^2 = C^2 = 0, C \bar C + \bar C C = 0$) fields
which are required to maintain the unitarity and ``quantum '' gauge invariance
(i.e. BRST invariance) together at any arbitrary order of perturbation
theory. The above Lagrangian density remains invariant
(up to a total derivative) under the following on-shell ($\Box C 
= \Box \bar C = 0$) nilpotent ($\tilde s_{(a)b}^2 = 0$) (anti-)BRST
symmetry transformations ($\tilde s_{(a)b}$) (with $\tilde s_{b} \tilde s_{ab}
+ \tilde s_{ab} \tilde s_{b} = 0$) as [31-33]
$$
\begin{array}{lcl}
\tilde s_{b} A_\mu &=& \partial_\mu C \;\qquad
\tilde s_{b} C = 0 \; \qquad \tilde s_{b} \bar C = - i (\partial \cdot A)
\nonumber\\
\tilde s_{ab} A_\mu &=& \partial_\mu \bar C\; \qquad
\tilde s_{ab} \bar C = 0\; \qquad 
\tilde s_{ab} C = + i (\partial \cdot A).
\end{array}\eqno(4.2)
$$
The same Lagrangian density (4.1) is also invariant (modulo some total
derivatives) under the following on-shell ($\Box C = \Box \bar C = 0$)
nilpotent ($\tilde s_{(a)d}^2 = 0$) (anti-)dual BRST symmetry transformations
$\tilde s_{(a)d}$ (with $\tilde s_{b} \tilde s_{ad} + \tilde s_{ad}
\tilde s_{d} = 0$) as [31-33]
$$
\begin{array}{lcl}
\tilde s_{d} A_\mu &=& - \varepsilon_{\mu\nu} \partial^\nu \bar C\; \qquad
\tilde s_{d} \bar C = 0\; \qquad \tilde s_{d} \bar C = - i E\;
\nonumber\\
\tilde s_{ad} A_\mu &=& - \varepsilon_{\mu\nu} \partial^\nu C\; \qquad
\tilde s_{ad} C = 0\; \qquad 
\tilde s_{ad} C = + i E.
\end{array}\eqno(4.3)
$$
The anti-commutator of these two symmetries 
(i.e. $\{ \tilde s_{b}, \tilde s_{d} \} = \{ \tilde s_{ab}, \tilde s_{ad} \}
= \tilde s_{w}) $ leads to the definition of
a bosonic symmetry ($\tilde s_{w}$). 
There exists a ghost scale symmetry transformation
for this theory where only the ghost fields undergo a scale transformation
and other fields remain untransformed. Together, the above symmetries obey 
an algebra that is reminiscent of the algebra obeyed
by the de Rham cohomology operators [31-36].

The kinetic energy- and gauge-fixing terms of the Lagrangian density (4.1)
can be linearized by introducing the auxiliary fields ${\cal B}$ and $B$ as 
[31-33]
$$
\begin{array}{lcl}
\tilde {\cal L}_{B} = {\cal B} E - \frac{1}{2} {\cal B}^2 
+ B (\partial \cdot A) + \frac{1}{2} B^2 - i \partial_\mu \bar C 
\partial^\mu C
\end{array}\eqno(4.4)
$$
which respects (modulo some total derivatives), the following off-shell
nilpotent ($ \tilde s_{(A)B}^2 = \tilde s_{(A)D}^2 = 0$) (anti-)BRST
($\tilde s_{(A)B}$) and (anti-)dual BRST 
($\tilde s_{(A)D}$) symmetry transformations
$$
\begin{array}{lcl}
\tilde s_{B} A_\mu &=& \partial_\mu C\; \qquad
\tilde s_{B} C = 0\; \qquad \tilde s_{B} \bar C = i B\;
\qquad \tilde s_{B} B = 0\; \qquad \tilde s_{B} {\cal B} = 0
\nonumber\\
\tilde s_{AB} A_\mu &=& \partial_\mu \bar C\; \qquad
\tilde s_{AB} \bar C = 0\; \qquad 
\tilde s_{AB} C = - i B\; \qquad \tilde s_{AB} B = 0\; \qquad
\tilde s_{AB} {\cal B} = 0 \nonumber\\
\tilde s_{D} A_\mu &=& - \varepsilon_{\mu\nu} \partial^\nu \bar C\; \quad
\tilde s_{D} \bar C = 0\; \qquad \tilde s_{D} C = - i {\cal B}\; \qquad
\tilde s_{D} {\cal B} = 0\; \qquad \tilde s_{D} B = 0
\nonumber\\
\tilde s_{AD} A_\mu &=& - \varepsilon_{\mu\nu} \partial^\nu C\; \quad
\tilde s_{AD} C = 0\; \quad 
\tilde s_{AD} \bar C = + i {\cal B}\; \quad \tilde s_{AD} {\cal B} = 0\;
\quad \tilde s_{AD} B = 0.
\end{array}\eqno(4.5)
$$
Here too, the anti-commutator $\{ \tilde s_{B}, \tilde s_{D} \} = 
\{\tilde s_{AB}, \tilde s_{AD} \} = \tilde s_{W}$ leads to the existence
of the off-shell version of a bosonic symmetry and a scale transformation
on the ghost fields defines a ghost symmetry. Thus, in the off-shell version
too, we have {\it six symmetries} in the theory. Corresponding to the equations
(3.7) and (3.8), we can write the expressions for the Lagrangian
density of the 2D free Abelian gauge theory as 
$$
\begin{array}{lcl}
\tilde {\cal L}_{B} &=& {\cal B} E - \frac{1}{2} {\cal B}^2 + X\; \qquad 
\tilde {\cal L}_{B} = 
B (\partial \cdot A) + \frac{1}{2} B^2 + Y \nonumber\\ 
X &=& \tilde s_{B} \bigl [ - i \bar C (\partial \cdot A + \frac{1}{2} B)
\bigr ] = 
\tilde s_{AB} \bigl [ + i C (\partial \cdot A + \frac{1}{2} B)
\bigr ] \nonumber\\
Y &=& \tilde s_{AD} \bigl [ - i \bar C (E -  \frac{1}{2} {\cal B})
\bigr ] = 
\tilde s_{D} \bigl [ + i C (E - \frac{1}{2} {\cal B})
\bigr ].
\end{array}\eqno(4.6)
$$
Moreover, modulo some total derivatives, the Lagrangian density (4.1) can be
expressed as the following sum of the (anti-)BRST and (anti-)co-BRST parts
$$
\begin{array}{lcl}
\tilde {\cal L}_{b} &=& \tilde s_{d} \bigl (\frac{i}{2} E C \bigr )
- \tilde s_{b} \bigl (\frac{i}{2} (\partial \cdot A) \bar C \bigr ) 
\equiv \{ \tilde Q_{d}, \frac{1}{2} E C \} - \{ \tilde Q_{b},
\frac{1}{2} (\partial \cdot A) \bar C \} \nonumber\\
\tilde {\cal L}_{b} &=& \tilde s_{ab} \bigl (\frac{i}{2} (\partial \cdot A)
C \bigr )
- \tilde s_{ad} \bigl (\frac{i}{2} E \bar C \bigr ) 
\equiv \{ \tilde Q_{ab}, \frac{1}{2} (\partial \cdot A) C \} 
- \{ \tilde Q_{ad},
\frac{1}{2} E \bar C \}
\end{array}\eqno(4.7)
$$
where conserved and on-shell nilpotent (anti-)co-BRST charges $\tilde Q_{(a)d}$ 
and  (anti-)BRST charges $\tilde Q_{(a)b}$ are the generators for
the on-shell nilpotent (co-)BRST symmetry transformations $\tilde s_{(a)d}$
and $\tilde s_{(a)b}$. The local expressions for these charges can be found
in [31-36]. We do not need these expressions for our present
discussions here. A close look at (4.6) and (4.7) 
implies that equation (4.7) for the 
Lagrangian density in (4.1) can be derived from (4.6) by exploiting the
equations of motion $ B = - (\partial \cdot A), {\cal B} = E$ and
substituting these in the expressions for $X$ and $Y$. The expression
for the Lagrangian density in (4.7) implies that the symmetric 
energy-momentum tensor $T_{\mu\nu}$
for the theory would also be the sum of (anti-)BRST and (anti-)co-BRST 
invariant parts [33,37]. As a consequence, both the Lagrangian density 
as well as the symmetric energy momentum tensor would be sum of 
(anti-)BRST and (anti-)co-BRST anti-commutators.
In  fact, the set of conserved charges $(Q_{b}, Q_{d}, Q_{w})$
(or $Q_{ad}, Q_{ab}, Q_{w})$ can be exploited to define the Hodge
decomposition of a state with ghost number $n$
(i.e. $ i Q_{g} |\psi>_{n} = n |\psi>_n$) in the quantum Hilbert space as
$$
\begin{array}{lcl}
|\psi>_{n} = |\omega>_{n} + Q_{b} |\theta>_{n-1} + Q_{d} |\chi>_{n+1}
\equiv  |\omega>_{n} + Q_{ad} |\theta>_{n-1} + Q_{ab} |\chi>_{n+1}
\end{array}
$$
where $|\omega>_n $ is the harmonic state, $Q_{b} |\theta>_{n-1}$ is the 
BRST exact state and $Q_{d} |\chi>_{n+1}$ is the co-BRST exact state. The
other states, in the above equation, can be defined similarly. The choice of
the harmonic state ($Q_{b} |phys> = 0, Q_{d} |phys> = 0$) to be the physical
state of the theory furnishes a proof for the topological nature of the
theory because there is no energy excitation 
(i.e. $<phys| \hat T_{00} |phys>^{\prime} = 0$) in the 
physical sector of the theory [33,38-41] as the symmetric energy momentum 
tensor is the sum of (co-)BRST anti-commutators
and the conserved, nilpotent and
{\it hermitian} (co-)BRST charges $Q_{(d)b}$ satisfy 
$Q_{b} |phys> = 0 = <phys| Q_b, Q_{d} |phys> = 0 = <phys | Q_d$.

Now we wish to dwell a bit on (3.7) and study some of the properties of
the 2-form antisymmetric Abelian gauge theory in 4D 
treating our discussions
on the one-form 2D free Abelian gauge theory as the backdrop. In fact,
the similarity  between (4.6) and (3.7) propel us to look into this aspect.
Exploiting the equations of motion of (2.7), it can be checked that, modulo
some total derivatives, $G$ and $F$ of (3.8) can be re-expressed as
$$
\begin{array}{lcl}
G \rightarrow G^{(0)} &=& - \frac{1}{4} C^\mu (\varepsilon_{\mu\nu\kappa\zeta}
\partial^\nu B^{\kappa\zeta}) - \frac{1}{2} (\partial \cdot \bar C) \beta
\nonumber\\
F \rightarrow F^{(0)} &=& + \frac{1}{2} \bar C^\mu 
(\partial^\nu B_{\nu\mu}) +\frac{1}{2} (\partial \cdot C) \bar \beta.
\end{array}\eqno(4.8)
$$
Now following the analogy with the 2D free Abelian one-form gauge theory
(discussed above), it can be checked that
$$
\begin{array}{lcl}
&&s_{d} (G^{(0)}) + s_{b} (F^{(0)}) = \tilde {\cal L}_{b}^{(0)} + \partial_\mu
[\; L^\mu\; ] \nonumber\\
&& L^\mu = \frac{1}{2}\; [ \; \partial^\mu (\beta\bar\beta)
- (\partial^\mu \bar C^\nu - \partial^\nu \bar C^\mu) C_{\nu}
- \bar C_\nu (\partial^\mu C^\nu - \partial^\nu C^\mu)\; \bigr ]
\end{array}\eqno(4.9)
$$
where the Lagrangian density $\tilde {\cal L}_{b}^{(0)}$ 
$$
\begin{array}{lcl}
\tilde {\cal L}_{b}^{(0)} &=& 
 \frac{1}{2} (\partial^\kappa B_{\kappa\mu} - \partial_\mu
\phi_{1}) (\partial_\sigma B^{\sigma\mu}) 
- \frac{1}{2} (\frac{1}{2} \varepsilon_{\mu\nu\kappa\zeta}
\partial^\nu B^{\kappa\zeta} - \partial_\mu \phi_{2}) 
(\frac{1}{2} \varepsilon^{\mu\xi\sigma\eta} \partial_\xi B_{\sigma\eta}) 
\nonumber\\
&-& \partial_\mu \bar \beta \partial^\mu \beta + (\partial_\mu \bar C_\nu
- \partial_\nu \bar C_\mu) (\partial^\mu C^\nu) - \frac{1}{2} (\partial \cdot
\bar C) (\partial \cdot C)
\end{array}\eqno(4.10)
$$
is different from the Lagrangian density (2.8) that respects the on-shell
nilpotent (anti-)BRST and (anti-)co-BRST symmetries. Even though the ghost
part of the above Lagrangian density is exactly same as in (2.8), there
are some terms in the bosonic part of (2.8) that are absent here. For
instance, it can  be checked that the  terms in (4.10), that contain the
auxiliary scalar fields $\phi_{1}, \phi_{2}$, are total derivatives. Thus,
they can be dropped from the total action as they do not influence
dynamics of the theory in any way. As a consequence, the above Lagrangian
density in (4.10) can be treated to be independent of the terms containing
fields $\phi_{1}$ and $\phi_{2}$ which are very much present in (2.8).
It will be noted that this feature of the 2-form 4D Abelian gauge theory
is drastically different from the corresponding 2D one-form Abelian gauge 
theory (which we have discussed above). The upshot of our whole discussion is
the fact that the Lagrangian density in (2.8) (that respects on-shell 
nilpotent (anti-)BRST and (anti-)co-BRST symmetries) {\it cannot} be
expressed as the sum of (anti-) BRST and (anti-)co-BRST anti-commutators
as is the case for the 2D free Abelian (cf (4.7)) and
self-interacting non-Abelian gauge theories [33,34,40,42].

Now let us take a different choice in place of $F^{(0)}$ and $G^{(0)}$
so that we can get the correct bosonic part of the Lagrangian density (2.8).
An interesting, such a choice, at our disposal is
$$
\begin{array}{lcl}
F^{(1)} = F^{(0)} + \frac{1}{2} (\partial \cdot \bar C) \phi_{1}\; \qquad
G^{(1)} = G^{(0)} - \frac{1}{2} (\partial \cdot C) \phi_{2}.
\end{array}\eqno(4.11)
$$
The variation of these expressions w.r.t. the on-shell nilpotent
(co-)BRST transformations can be written explicitly as
$$
\begin{array}{lcl}
s_{b} (F^{(1)}) &=& 
\frac{1}{2} (\partial^\nu B_{\nu\mu} - \partial_\mu
\phi_{1}) (\partial_\sigma B^{\sigma\mu} - \partial^\mu \phi_{1}) 
- \frac{1}{2} \partial_\mu \bar \beta \partial^\mu \beta \nonumber\\ 
&+& \frac{1}{2} (\partial_\mu \bar C_\nu
- \partial_\nu \bar C_\mu) (\partial^\mu C^\nu) - \frac{1}{2} (\partial \cdot
\bar C) (\partial \cdot C) + \partial_\mu M^\mu \nonumber\\
s_{d} (G^{(1)}) &=& 
- \frac{1}{2} (\frac{1}{2} \varepsilon_{\mu\nu\kappa\zeta}
\partial^\nu B^{\kappa\zeta} - \partial_\mu
\phi_{2}) (\frac{1}{2} \varepsilon^{\mu\xi\sigma\eta}
\partial_\xi B_{\sigma\eta} - \partial^\mu \phi_{2}) \nonumber\\
&-& \frac{1}{2} \partial_\mu \bar \beta \partial^\mu \beta 
+ \frac{1}{2} (\partial_\mu \bar C_\nu
- \partial_\nu \bar C_\mu) (\partial^\mu C^\nu) - \frac{1}{2} (\partial \cdot
\bar C) (\partial \cdot C) + \partial_\mu N^\mu 
\end{array}\eqno(4.12)
$$
where the expressions for the $M^\mu$ and $N^\mu$ in terms of the local
fields are
$$
\begin{array}{lcl}
M^\mu &=& \frac{1}{2}\;
\bigl [\;
\bar\beta \partial^\mu \beta - \bar C_\nu (\partial^\mu C^\nu - \partial^\nu
C^\mu) + (\partial_\nu B^{\nu\mu} - \partial^\mu \phi_{1}) \phi_{1}\; \bigr ]
\nonumber\\
N^\mu &=& \frac{1}{2}\;
\bigl [\;
\beta \partial^\mu \bar \beta - (\partial^\mu \bar C^\nu - \partial^\nu
\bar C^\mu) C_\nu 
+ (\frac{1}{2} \varepsilon^{\mu\nu\kappa\zeta} \partial_\nu B_{\kappa\zeta}
- \partial^\mu \phi_{2}) \phi_{2} \;\bigr ].
\end{array}\eqno(4.13)
$$
The sum of (4.11) and (4.12) yields, modulo some total derivatives,
the Lagrangian density of (2.8) {\it but for the last term} as 
$$
\begin{array}{lcl}
{\cal L}_{b}^{(0)} &=& \partial_\mu ( P^\mu ) + 
 \frac{1}{2} (\partial^\nu B_{\nu\mu} - \partial_\mu
\phi_{1}) (\partial_\sigma B^{\sigma\mu} - \partial^\mu \phi_{1}) \nonumber\\
&-& \frac{1}{2} (\frac{1}{2} \varepsilon_{\mu\nu\kappa\zeta}
\partial^\nu B^{\kappa\zeta} - \partial_\mu \phi_{2}) 
(\frac{1}{2} \varepsilon^{\mu\xi\sigma\eta} \partial_\xi B_{\sigma\eta} -
\partial^\mu \phi_{2}) \nonumber\\
&-& \partial_\mu \bar \beta \partial^\mu \beta + (\partial_\mu \bar C_\nu
- \partial_\nu \bar C_\mu) (\partial^\mu C^\nu) - 
({\bf \frac{1}{2} + \frac{1}{2}}) (\partial \cdot \bar C) (\partial \cdot C)
\end{array}\eqno(4.14)
$$
where the local expression for $P^\mu$ is
$$
\begin{array}{lcl}
P^\mu &=& \frac{1}{2}\;
\bigl [\;
 \partial^\mu (\bar\beta\beta) - \bar C_\nu (\partial^\mu C^\nu - \partial^\nu
C^\mu) + (\partial_\nu B^{\nu\mu} - \partial^\mu \phi_{1}) \phi_{1}
\nonumber\\
&-& (\partial^\mu \bar C^\nu - \partial^\nu
\bar C^\mu) C_\nu 
+ (\frac{1}{2} \varepsilon^{\mu\nu\kappa\zeta} \partial_\nu B_{\kappa\zeta}
- \partial^\mu \phi_{2}) \phi_{2} \;\bigr ].
\end{array}\eqno(4.15)
$$
Thus, we note that {\it except for an extra factor of half in the last term},
the Lagrangian density for the 4D 2-form Abelian gauge theory can be written
as the sum of (anti-)BRST and (anti-)co-BRST anti-commutators. It is evident
now that the symmetric energy momentum for the theory will also not be able
to be expressed as the sum of two anti-commutators. Hence, the free 2-form
Abelian gauge theory in 4D is {\it not} an example of an
exact TFT but (i) it is
a field theoretical model for the Hodge theory in 4D, and (ii) its properties
are such that it can be treated as an example of a field theoretical model
for the q-TFT.\\

\noindent
{\bf 5 Conclusions}\\

\noindent
In the present investigation,
we studied the free 4D Abelian 2-form (antisymmetric) gauge theory from the
point of view of (i) its cohomological properties, and
(ii) its quasi-topological nature.  In the first part, we 
invoked some of the results of our earlier work [25]
because these inferences and issues were pertinent 
and complementary to our present work. We found that the above
 theory provides a tractable field theoretical model for the Hodge theory
in the physical four $(3 + 1)$-dimensions of spacetime. The symmetries of 
the Lagrangian density correspond to all the
de Rham cohomological operators of differential geometry. In fact,
there are six local, covariant and continuous symmetries in
the theory of which four ($s_{(A)B},s_{(A)D}$) symmetries are nilpotent
($s_{(A)B}^2 = s_{(A)D}^2 = 0$) of order two. The latter symmetries 
are very interesting and they play a pivotal role in expressing the 
``topological'' invariants of the theory which obey a certain specific
type of recursion relations. Moreover, these symmetries and their
anticommutator (which corresponds to a bosonic symmetry) obey an 
algebra that is also respected by the corresponding conserved Noether charges
$Q_{r} (r = AB, B, AD, D, W)$ for the off-shell version of the symmetries [25]
$$
\begin{array}{lcl}
Q_{(A)B}^2 &=& 0 \; \qquad Q_{(A)D}^2 = 0 \; \qquad \{ Q_{B}, Q_{D} \} = Q_{W}
= \{Q_{AB}, Q_{AD} \} 
\nonumber\\
\{ Q_{AB}, Q_{B} \} &=& 0 \; \qquad \{ Q_{D}, Q_{AD} \} = 0 \; \qquad
[ Q_{W}, Q_{r} ] = 0. 
\end{array}\eqno(5.1)
$$
The above algebra is reminiscent of the algebra obeyed by 
$(d, \delta, \Delta)$. It will be noted that the nilpotent pair $(d,\delta)$
together can not be identified with $(Q_{B}, Q_{AB})$
or $(Q_{D},Q_{AD})$ because the latter pairs of generators  anti-commute
with each-other in their set whereas the former pair $(d,\delta)$ does not.
The algebra among the cohomological operators can be succinctly expressed as 
$$
\begin{array}{lcl}
&& d^2 = 0 \qquad  \delta^2 = 0 \qquad \Delta =
(d + \delta)^ 2 = d \delta + \delta d \nonumber\\
&& [ \Delta , d ] = 0 \qquad  [\Delta , \delta ] = 0 \qquad
\Delta = \{ d, \delta \} \neq 0.
\end{array}\eqno(5.2)
$$
The algebra between conserved charges and the ghost charge $Q_{g}$
$$
\begin{array}{lcl}
i [ Q_{g}, Q_{B(AD)} ] = + Q_{B(AD)}  \qquad 
i [ Q_{g}, Q_{D(AB)} ] = - Q_{D(AB)} \qquad
i [ Q_{g}, Q_{W} ] = 0  
\end{array}\eqno(5.3)
$$
is such that any arbitrary state $|\psi>_{n}$
in the the quantum Hilbert space of states
with ghost number $n$ (i.e. $i Q_{g} |\psi>_{n} = n |\psi>_{n}$) would
imply the following (see, e.g., [25])
$$
\begin{array}{lcl}
i Q_{g} Q_{B(AD)} | \psi >_{n} &=& (n + 1) Q_{B(AD)} |\psi>_{n} \;\nonumber\\
i Q_{g} Q_{D(AB)} | \psi >_{n} &=& (n - 1) Q_{D(AB)} |\psi>_{n} \;
\nonumber\\
i Q_{g} Q_{W}   \; | \psi >_{n}  &=& n \;    Q_{W} \;  |\psi>_{n}.
\end{array}\eqno(5.4)
$$   
This shows that the ghost numbers for the 
states $Q_{B(AD)} |\psi>_{n}$, $Q_{D(AB)} |\psi>_{n}$ and 
$Q_{W} |\psi>_{n}$ 
in the quantum Hilbert space 
are $ (n + 1), (n - 1)$ and $ n $, respectively. This feature is same as
the result of the  application of the cohomological operators 
($d,\delta,\Delta$)
on a differential form $f_{n}$ of degree $n$. This analogy is valid
due to the fact that the forms 
$d f_{n}, \delta f_{n}$ and $\Delta f_{n}$ possess
the degree $n + 1, n - 1$ and $n$, respectively. This discussion, allows
us to find a mapping between cohomological operators and conserved charges
as: $d \leftrightarrow Q_{B(AD)}, \delta \leftrightarrow Q_{D(AB)},
\Delta \leftrightarrow Q_{W}$. This establishes the fact that 4D free
Abelian 2-form gauge theory is a field theoretical model for the Hodge
theory because the discrete transformations in (2.4) correspond to
the Hodge duality $*$ operation of the differential geometry.

Now we wish to state some of the key results of the second part of our
investigation.
We focussed on some of the distinguished properties of the 4D 2-form 
Abelian gauge theory {\it vis-{\`a}-vis} 2D topological Abelian 
(as well as self-interacting non-Abelian) one-form gauge theory. We
observed some common features and a few 
striking differences between these theories.
The overlapping similarities are: (i)
the form of the Lagrangian densities of both the theories
can be expressed in the same fashion (cf eqns (3.7) and (4.6)).
It should be emphasized, however, that the above
statement is found to be 
true only when we exploit the {\it off-shell} nilpotent version
of the (anti-)BRST and (anti-)co-BRST symmetries. (ii) We could derive
``topological invariant'' quantities on the 4D manifold that
are similar to the topological invariants of the 2D one-form
(non-)Abelian gauge theories defined on the 2D manifold [31-34,38-43].
(iii) These invariants obey the same kind of recursion relations as
invariants of topological field theories. (iv) Both the theories
(i.e. 4D 2-form and 2D one-form free Abelian gauge theories) provide a set of 
tractable field theoretical models for the Hodge theory. (v) The Lagrangian
densities of both the theories respect six local, covariant and continuous
symmetries. Now, we dwell a bit on the  key differences
between these two theories. These are: (i)
as far as the form and appearance of the Lagrangian densities 
(and symmetric energy-momentum tensor) are concerned,
there exists a slight difference between the two theories when they are
expressed in terms of the {\it on-shell} version of the nilpotent symmetries.
It is clear from (4.7) that the Lagrangian density 
(so the symmetric energy momentum tensor) for the 2D free Abelian
(and self-interacting non-Abelian) gauge theories can be expressed, modulo
some total derivatives, as the sum of (anti-)BRST and (anti-)co-BRST
anti-commutators. This is not the case with the 4D 2-form Abelian
gauge theory as can be seen from (4.14). In fact, the Lagrangian density
(and therefore the symmetric energy-momentum tensor) for the latter
theory {\it cannot} be exactly expressed as the sum
of (anti-)BRST and (anti-)co-BRST anti-commutators. {\it The two theories
mathematically resemble very much but for a factor of half}.
(ii) The free 2D (non-)Abelian gauge theories, defined on the 2D manifold
without a non-trivial topology at the boundary, are topological in nature 
but the same is not true for the 4D 2-form Abelian gauge theory. In fact,
the counting of degrees of freedom also shows it. However, some of the
striking similarities between these two theories propel us to conclude that
4D 2-form gauge theory is possibly
an example of the field theoretical models for the q-TFT.
(iii) The topological invariants $I_{k}(\tilde I_{k})$ and
$J_{k}(\tilde J_{k})$ for the 2D (non-)Abelian theories are defined 
on the 2D closed Riemann surface (i.e. the Eauclidean version [51] of 
the 2D manifold) for all the possible values of $ k = 0, 1, 2$. The analogous
thing does not happen for the present 4D 2-form gauge theory because
{\it all} the invariants do not exist for all the possible values
of $k = 0, 1, 2, 3, 4$. At least, in our present investigation, we are
unable to provide a logical reason behind this discrepancy. Perhaps, this
difference too corroborates the non-topological nature of 4D
2-form gauge theory.
(iv) There exist four sets of topological invariants w.r.t. (anti-)BRST
and (anti-)co-BRST charges of the 2D theories because of the fact that
there is a single set of fermionic scalar (anti-)ghost fields $(\bar C)C$.
This is not the case with 4D 2-form Abelian gauge theory. In fact,
for the latter theory, we have a single set of fermionic vector (anti-)ghost 
fields $(\bar C_\mu)C_\mu$, a single set of fermionic scalar (anti-)ghost
fields $(\rho)\lambda$ and a single set of bosonic (anti-)ghost fields
$(\bar\beta)\beta$. Therefore, four sets of invariants exist for each
of these sets w.r.t. (anti-)BRST and (anti-)co-BRST charges which makes the
total number of invariants to be twelve. These have been computed in
section 3 of our present paper.\\

\noindent
{\bf Acknowledgements}\\

\noindent
Constructive, clarifying and critical comments by the referees' are
gratefully acknowledged. Thanks are also due to E. Harikumar and 
A. Lahiri for taking interest in this work. \\

\end{document}